\documentclass[prb,twocolumn,preprintnumbers,amsmath,amssymb,superscriptaddress]{revtex4}

\usepackage{graphicx}
\usepackage{dcolumn}
\usepackage{bm}
\usepackage{color}

\begin{document}

\title{Structural investigation of the bilayer iridate Sr$_{3}$Ir$_{2}$O$_{7}$}

\author{Tom Hogan}
 \affiliation{Department of Physics, Boston College, Chestnut Hill, Massachusetts 02467, USA}
 \affiliation{Materials Department, University of California, Santa Barbara, California 93106-5050, USA}

\author{Lars Bjaalie}
 \affiliation{Materials Department, University of California, Santa Barbara, California 93106, USA}

\author{Liuyan Zhao}
 \affiliation{Department of Physics, California Institute of Technology, Pasadena, California 91125, USA}
 \affiliation{Institute for Quantum Information and Matter, California Institute of Technology, Pasadena, California 91125, USA}

\author{Carina Belvin}
\affiliation{Wellesley College, Wellesley, Massachusetts 02481, USA}

\author{Xiaoping Wang}
\affiliation{Chemical and Engineering Materials Division, Oak Ridge National Laboratory, Oak Ridge, Tennessee 37831, USA}

\author{Chris G. Van de Walle}
\affiliation{Materials Department, University of California, Santa Barbara, California 93106-5050, USA}

\author{David Hsieh}
 \affiliation{Department of Physics, California Institute of Technology, Pasadena, California 91125, USA}
 \affiliation{Institute for Quantum Information and Matter, California Institute of Technology, Pasadena, California 91125, USA}

\author{Stephen D. Wilson}
 \email{stephendwilson@engineering.ucsb.edu}
 \affiliation{Materials Department, University of California, Santa Barbara, California 93106-5050, USA}
 
\begin{abstract}

A complete structural solution of the bilayer iridate compound Sr$_{3}$Ir$_{2}$O$_{7}$ presently remains outstanding. Previously reported structures for this compound vary and all fail to explain weak structural violations observed in neutron scattering measurements as well as the presence of a net ferromagnetic moment in the basal plane.  In this paper, we present single crystal neutron diffraction and rotational anisotropy second harmonic generation measurements unveiling a lower, monoclinic symmetry inherent to Sr$_{3}$Ir$_{2}$O$_{7}$.   Combined with density functional theory, our measurements identify the correct structural space group as No.\ 15 (\textit{C2/c}) and provide clarity regarding the local symmetry of Ir$^{4+}$ cations within this spin-orbit Mott material.
\end{abstract}

\maketitle

\section{Introduction}

The spin-orbit Mott (SOM) state as realized in the Ruddlesden-Popper (RP) strontium iridates (Sr$_{n+1}$Ir$_{n}$O$_{3n+1}$) remains of considerable interest due to the unique physics arising from the confluence of comparable strong spin-orbit coupling and electron-electron interactions in the presence of a strong cubic crystal field.\cite{Moon2008, BJKim2008, BJKim2009}  Subtleties in the lattice structures of these systems are of critical importance in determining both the spin-orbital content of the ground state wave function,\cite{BJKim2009} the resulting magnetic ground state,\cite{Jackeli2009} and details within the electronic band structure.\cite{Carter2013}  In particular, for the $n=2$ member of the RP series (Sr$_3$Ir$_2$O$_7$) both density functional theory (DFT) \cite{King2013,Grioni2014,Liu2014} and models of magnetic order \cite{PhysRevLett.109.037204} are sensitive to subtleties in the local crystal fields at Ir sites.

The structure of Sr$_{3}$Ir$_{2}$O$_{7}$ was originally reported as a distorted bilayer perovskite variant described by the tetragonal space group \textit{I4/mmm} (No.\ 139) with unit cell dimensions \textit{a} = 3.896 \AA, \textit{c} = 20.879 \AA.\cite{Subramanian94}  In initial studies, evidence of an in-plane rotation (along the unique tetragonal \textit{c} axis) of the octahedral cages surrounding the Ir atoms was observed via x-ray scattering; however this was refined within a disordered \textit{I4/mmm} model, where the phasing of the octahedral rotations from site to site is random.  Other subsequent x-ray investigations have also observed incoherent rotations among the oxygen octahedra\cite{Boseggia2012}---potentially arising from compositional disorder.  

A separate x-ray study put forward an orthorhombic unit cell (\textit{a} = 5.522 \AA, \textit{b} = 5.521 \AA, \textit{c} = 20.917 \AA) indexed with an improper Hermann-Mauguin symbol \textit{Bbca}\cite{Cao2002} (space group No.\ 68) and modeled using coherent intralayer counter-rotations of neighboring octahedra along the c-axis.  Additional support for this orthorhombic model with coherent octahedral rotations was provided by a transmission electron microcopy (TEM) study of the reflection conditions observed in electron diffraction patterns.\cite{Matsuhata2004}  Here the reciprocal lattices of both \textit{Bbcb}- and \textit{Acaa}-type structures, each an alternate setting of the \textit{Ccce} space group (No.\ 68), were superimposed to match the observed patterns---demonstrating that this system is prone to twinning.  This TEM-derived structural solution parallels that of the analogous ruthenate compound Sr$_{3}$Ru$_{2}$O$_{7}$,\cite{Shaked2000} where all modes of octahedral rotation about their symmetry axes were considered as candidate structures.  

Even when accounting for the coherent phasing of octahedral rotations, \textit{Ccce} still fails to completely describe some subtle aspects of Sr$_{3}$Ir$_{2}$O$_{7}$'s lattice structure.  Neutron diffraction measurements\cite{Dhital2012} have observed peaks in the [H 0 L] zone (\textit{Bbcb} setting) that violate the reflection condition H,L = 2\textit{n} imposed by the space group.\cite{IntTables}  These weak violations were later confirmed to be of structural origin via a polarized neutron scattering study\cite{Dhital2014} and are reminiscent of those observed within the $n = 1$ system Sr$_2$IrO$_4$ (also via neutron diffraction).\cite{Feng2013, Dhital2013} While in Sr$_2$IrO$_4$ two unique Ir environments were ultimately refined,\cite{Torch2015,Feng2015}  the origin of the Bragg violations in Sr$_{3}$Ir$_{2}$O$_{7}$ and their implication for the lattice structure remains an open question.

In this paper, we utilize single crystal neutron scattering and rotational anisotropy second harmonic generation (RA-SHG) measurements, to resolve the structure of Sr$_{3}$Ir$_{2}$O$_{7}$.  The point group for the lattice is constrained via RA-SHG measurements to be either $4/m$ or $2/m$ or one of even lower symmetry, which when combined with single crystal neutron data identifies the monoclinic space group \textit{C2/c} as the correct structural symmetry. Density functional theory calculations were used to guide the space group search and identify the most energetically favored mode of lattice distortion.  In addition to in-plane rotations, oxygen octahedra in this new lower symmetry can tilt off-axis, suggesting that the anomalous, weak in-plane ferromagnetism of Sr$_{3}$Ir$_{2}$O$_{7}$ originates from these tilts combined with strong spin-lattice coupling.  Using probes sensitive to both oxygen sites as well as point group symmetry, our measurements ultimately provide a foundation for understanding the further structural distortions observed in this system during metallization via pressure \cite{Zhao2014} and electron substitution.\cite{Hogan2015}

\begin{figure}
\includegraphics[trim = 50mm 0mm 0mm 0mm, clip, scale=0.5]{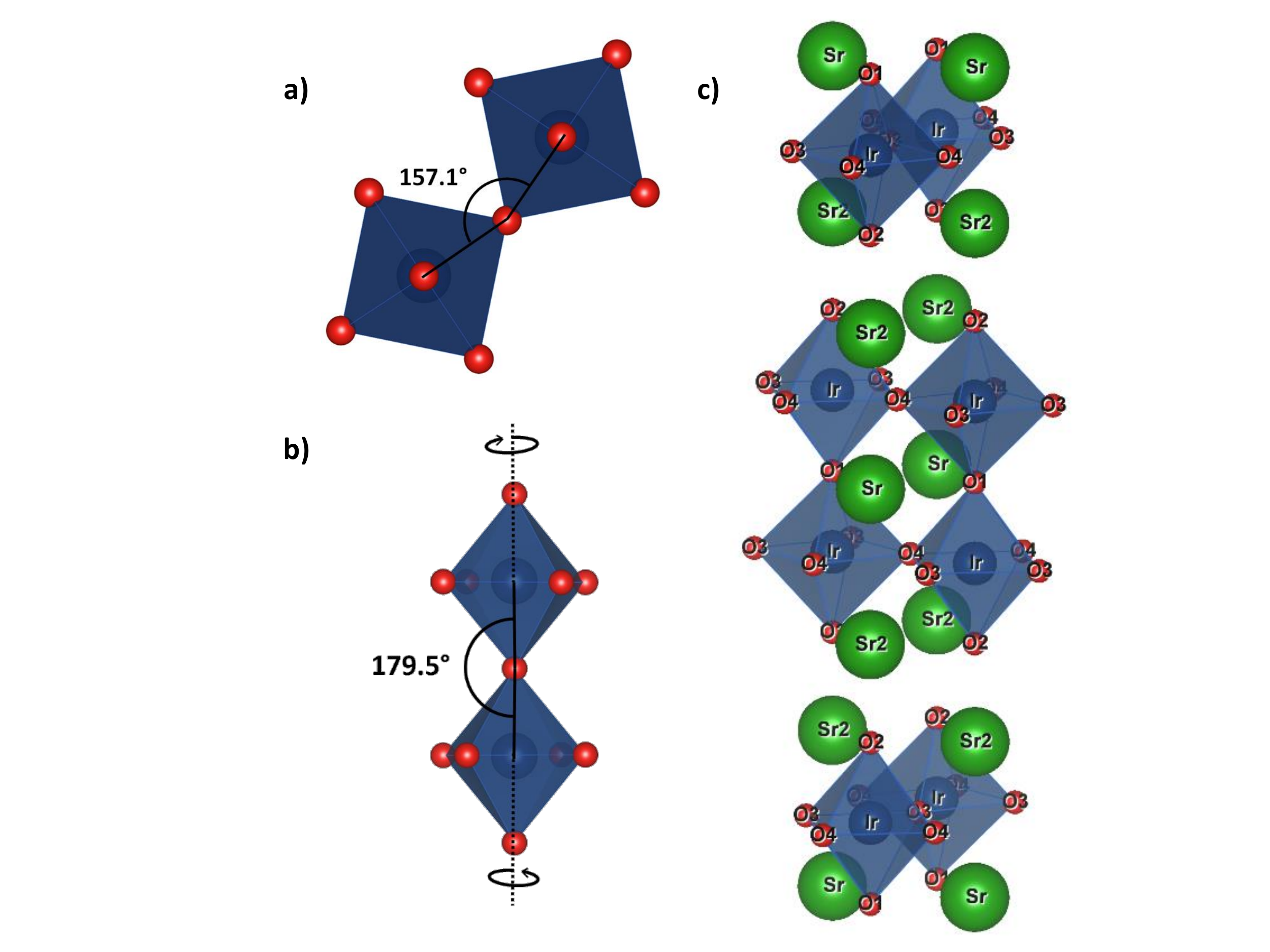}
\caption{\label{fig:unitcell} Refined lattice structure of Sr$_3$Ir$_2$O$_7$.  Red atoms denote oxygen sites, blue atoms within octahedra denote iridium sites, and green atoms denote strontium sites.  (a) The in-plane oxygen octahedral rotations as viewed along the \textit{a}-axis.  (b) Out-of plane oxygen octahedral tilt mode, viewed along the \textit{b}-axis.  c) Off-axis view of the chemical unit cell.  Atom positions are those of the $T = 300$ K refinement to the \textit{C2/c} space group.}  
\end{figure}

\section{Experimental and Computational Details}

The method for growing the  Sr$_3$Ir$_2$O$_7$ single crystals used in our study is reported elsewhere.\cite{Hogan2015}  Neutron scattering measurements were conducted by measuring a small single crystal (mass 5.6 mg, dimensions 1.25 mm $\times$ 1.13 mm $\times$ 0.1 mm) in the time-of-flight single crystal Laue diffractometer TOPAZ at Oak Ridge National Laboratory. Sample orientations were optimized with the \textsc{crystalplan} software\cite{crystalplan} for an estimated coverage of 99.8\% of equivalent reflections for the nominal orthorhombic cell for this system.  Reduction of the raw data including Lorentz corrections, absorption, time-of-flight spectrum, and detector efficiency corrections were carried out with \textsc{anvred3}.\cite{anvred} The raw peaks were integrated using a 3-D ellipsoidal routine,\cite{ellipint} and the reduced dataset was refined using \textsc{shelxl}.\cite{shelx}

To guide the search for the correct space group, density functional theory was employed to compare several potential solutions.  Of the resulting relaxed structural geometries, those that did not display the correct antiferromagnetic ordering of the Ir 5d moments in the simulated structures could be discarded.  Structures which remained served as initial conditions for refinement to the neutron diffraction dataset.  The calculations were performed using the projector augmented-wave method \cite{PAW} in the Vienna Ab initio Simulation Package (VASP). \cite{KresseVASP1, KresseVASP2} The PBE functional \cite{PBE} was used with a screened on-site Coulomb repulsion parameter $U$ of 2 eV on the Ir \textit{d} orbitals, and spin-orbit coupling (SOC) taken into account. Taken together, the $+U$ interaction and SOC open up a Mott-Hubbard gap within the Ir $5d$ band.  Since the material is layered, Van der Waals interactions are accounted for via Becke-Johnson damping,\cite{BJ2005} as implemented in VASP. For the 48-atom unit cell, a plane-wave cutoff of 500 eV and 4 $\times$ 4 $\times$ 1 $ \Gamma$-centered k-point mesh were used.

Determining the point group of Sr$_{3}$Ir$_{2}$O$_{7}$ was accomplished through the use of RA-SHG techniques.  These data were acquired using the rotating scattering plane based technique described in Ref.\ [\onlinecite{Torch2014}] from cleaved surfaces of Sr$_{3}$Ir$_{2}$O$_{7}$ with the long cell axis parallel to the surface normal. Incident light was provided by a Ti:sapph regenerative amplifier (800 nm center wavelength, 60 fs pulse duration, 10 kHz repetition rate) and focused to a spot size less than 100 $\mu$m on the crystal with a fluence less than 1 mJ/cm\textsuperscript{2}. The linear polarization of the incident (in) and reflected (out) light was selected to be either in (P) or out of (S) the scattering plane. The orientation of the two in-plane crystallographic axes were determined independently by x-ray Laue diffraction.

\begin{center}
\centering
\begin{figure*}[th]
\includegraphics[scale=0.65]{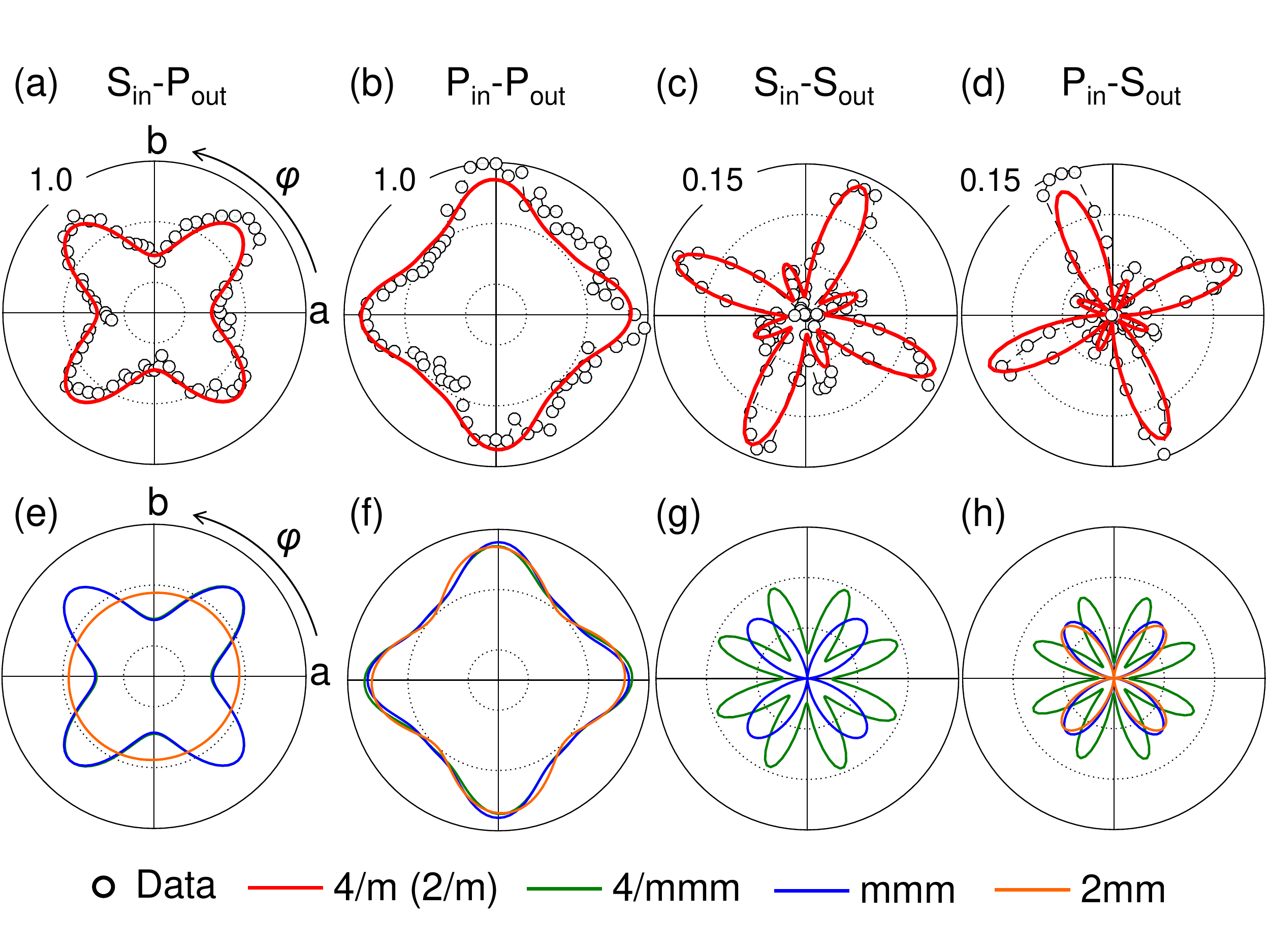}
\caption{\label{fig:SHGplot} RA-SHG patterns (open circles) from Sr$_{3}$Ir$_{2}$O$_{7}$ acquired under (a) S$_{in}$-P$_{out}$, (b) P$_{in}$-P$_{out}$, (c) S$_{in}$-S$_{out}$ and (d) P$_{in}$-S$_{out}$ polarization geometries at \textit{T} = 295 K. The intensities of all patterns are normalized against the PP trace. Red lines overlaid on the data are best fits to bulk electric quadrupole induced RA-SHG calculated using either of the centrosymmetric 4/m or 2/m point groups. The bottom row (e)\textendash(h) shows the corresponding best fits to bulk electric-quadrupole induced RA-SHG from the centrosymmetric 4/mmm (green) and mmm (blue) point groups as well as to bulk electric-dipole induced RA-SHG from the non-centrosymmetric 2mm point group (orange). Responses that are absent in the plots are forbidden by symmetry.}
\end{figure*}
\end{center}

\section{Neutron Diffraction Results}

We first examine the refined results from single crystal diffraction data collected from Sr$_{3}$Ir$_{2}$O$_{7}$.  Diffraction data were collected and patterns refined both at 100 K and 300 K.  As a starting procedure for refinement, the undistorted tetragonal parent \textit{I4/mmm} structure was transformed into candidate lower-symmetry space groups\cite{Bilbao1,Bilbao2,Bilbao3,subgroup} corresponding to pure rotations about axes of high symmetry.\cite{Shaked2000}  Distortions of the octahedra were introduced using the relaxed structures from DFT calculations as initial positions.  These structures were then refined to the single crystal neutron diffraction data at $T = 300$ K to evaluate which set of octahedral rotations most faithfully reproduced the observed pattern. Neutron scattering as a probe is particularly well-suited to this task as, compared to synchrotron radiation, it is more sensitive to the oxygen comprising the octahedra.  Similar to Sr$_{3}$Ir$_{2}$O$_{7}$'s ruthenate analog, all tetragonal subgroups rendered poor fits and the two best models which emerged were refined within space groups No.\ 68 (\textit{Ccce}) and No.\ 15 (\textit{C2/c}).  

The former group \textit{Ccce} is consistent with the previously reported structure,\cite{Cao2002,Matsuhata2004} and represents the coordinated rotation of in-plane octahedra with the rotational sense about the long (here, \textit{b}) axis for intralayer neighboring cages being opposite.  However this structure fails to account for the weak Bragg violations ($>10^3$ times weaker than primary peaks) known to exist in this system, implying that there may be a further distortion into a lower symmetry.  Alternatively, the latter \textit{C2/c} monoclinic group, related to \textit{Ccce} by the transformation matrix -\textit{b}+$\frac{1}{4}$, \textit{a}+$\frac{1}{4}$, \textit{c},  can account for these violations and represents a combination of the previous in-plane octahedral rotation with an additional octahedral tilt mode,\cite{Shaked2000} as seen projected separately in panels (a) and (b) of Fig.\ \ref{fig:unitcell}.  Taken blindly however, the \textit{R1} values resulting from refinement of the time-of-flight data alone were comparable for each candidate group, mandating further constraints to determine the correct solution.

\begin{table*}
\caption{\label{tab:coords} Results of refinement of \textit{T} = 300 K neutron diffraction data to the \textit{C2/c} model.  Wyckoff site labels, relative atomic coordinates and anisotropic displacement factor matrices $U_{ij}$ are included. \textit{R1} = 0.057}
\begin{ruledtabular}
\begin{tabular}{lllllll}
 Atom &Wyckoff Site &x &y &z  &U$_{11}$ \\
 \hline
   Ir     &8f     &0.59755(4)   &0.7495(5)     &0.7500(4) &0.00670\\
   Sr(1)  &4e     &0.500000     &0.2489(11)    &0.750000  &0.01010\\
   Sr(2)  &8f     &0.68747(7)   &0.7507(9)     &0.2494(5) &0.00930\\
   O(1)   &4e     &0.500000     &0.7480(14)    &0.750000  &0.00460\\
   O(2)   &8f     &0.69414(8)   &0.7487(9)     &0.7496(7) &0.00400\\
   O(3)   &8f     &0.09674(14)  &0.4499(9)     &0.4490(5) &0.01450\\
   O(4)   &8f     &0.09610(14)  &0.9488(9)     &0.5507(5) &0.01650\\
\end{tabular}

\begin{tabular}{lllllll}
 Atom  &U$_{22}$ &U$_{33}$ &U$_{12}$ &U$_{13}$ &U$_{23}$\\
 \hline
   Ir      &0.00800  &0.00210 &-0.00030 &-0.00170 &-0.00040 \\
   Sr(1)   &0.00700  &0.01200 & 0.00000 &-0.00100 & 0.00000 \\
   Sr(2)   &0.01200  &0.01100 & 0.00190 &-0.00190 &-0.00170 \\
   O(1)    &0.02100  &0.02200 & 0.00000 &-0.00230 & 0.00000 \\
   O(2)    &0.00700  &0.02100 & 0.00130 &-0.00180 &-0.00050 \\
   O(3)    &0.01120  &0.00700 &-0.00010 &-0.00040 & 0.00290 \\
   O(4)    &0.01040  &0.00840 & 0.00060 &-0.00100 &-0.00270 \\
	
\end{tabular}

\end{ruledtabular}
\end{table*}


\begin{table*}
\caption{\label{tab:coords100} Results of refinement of \textit{T} = 100 K neutron diffraction data to the \textit{C2/c} model.  Wyckoff site labels, relative atomic coordinates and anisotropic displacement factor matrices $U_{ij}$ are included.  \textit{R1} = 0.059}
\begin{ruledtabular}
\begin{tabular}{lllllll}
 Atom &Wyckoff Site  &x &y &z   &U$_{11}$\\
 \hline
   Ir	      &8f     &0.59754(4)   &0.7502(11)    &0.7504(2)  &0.00720\\  
   Sr(1)    &4e      &0.500000     &0.2490(2)     &0.750000  &0.00890 \\ 
   Sr(2)    &8f      &0.68756(7)   &0.7491(16)    &0.2505(4)  &0.01070\\  
   O(1)	    &4e      &0.500000     &0.7481(16)    &0.750000   &0.00700\\  
   O(2)	    &8f      &0.69432(8)   &0.7490(10)    &0.7502(5)  &0.00510\\  
   O(3)	    &8f      &0.09592(14)  &0.4460(9)     &0.4465(5) &0.01130 \\  
   O(4)	    &8f      &0.09688(16)  &0.9476(10)    &0.5525(7)  &0.01140\\
\end{tabular}

\begin{tabular}{lllllll}
 Atom  &U$_{22}$ &U$_{33}$ &U$_{12}$ &U$_{13}$ &U$_{23}$\\
 \hline
   Ir	      &0.00930  &0.00240  &0.00140 & 0.00060 &-0.00060 \\
   Sr(1)	  &0.00300  &0.00800  &0.00000 & 0.00310 & 0.00000 \\
   Sr(2)	  &0.00300  &0.01100  &0.00000 & 0.00610 &-0.00260 \\
   O(1)	    &0.00400  &0.01900  &0.00000 & 0.00800 & 0.00000 \\
   O(2)	    &0.00700  &0.01000  &0.00100 & 0.00180 & 0.00320 \\
   O(3)	    &0.00620  &0.00390  &0.00020 &-0.00080 &-0.00040 \\
   O(4)	    &0.01160  &0.00790  &0.00000 &-0.00060 &-0.00460 \\
	
\end{tabular}
\end{ruledtabular}
\end{table*}

\section{Rotational Anisotropy Second Harmonic Generation Results}

To further distinguish between the two possible solutions, rotational anisotropy second harmonic generation measurements were taken.  RA-SHG is a technique capable of directly determining the crystallographic point group symmetry of a material. In these experiments, light of frequency $\omega$ is obliquely incident on the surface of a crystal and the intensity of light reflected at 2$\omega$ is measured as a function of the angle ($\phi$) between the scattering plane and some in-plane crystalline axis.\cite{Torch2014} By performing these measurements using different combinations of incident and outgoing light polarization, the entire nonlinear optical susceptibility tensor can be determined, which embeds all the point group symmetries of the material. This technique has recently been used to help identify subtle structural distortions in Sr$_{2}$IrO$_{4}$\cite{Torch2015,Zhao2015} that lower the point group symmetry from 4/mmm to 4/m.

Fig.\ \ref{fig:SHGplot} shows RA-SHG patterns from Sr$_{3}$Ir$_{2}$O$_{7}$ acquired under all four linear polarization geometries at room temperature ($>T_N$). The bottom row shows best fits to calculations based on the three crystallographic point groups that have been proposed for Sr$_{3}$Ir$_{2}$O$_{7}$: tetragonal 4/mmm (\textit{I4/mmm}),\cite{Subramanian94} orthorhombic mmm (\textit{Bbcb}),\cite{Cao2002,Matsuhata2004} and orthorhombic 2mm (\textit{Bb}2$_{1}$\textit{m}).\cite{Dhital2014} For the non-centrosymmetric 2mm point group, we assume the dominant contribution to SHG to be of bulk electric-dipole origin. For the centrosymmetric 4/mmm and mmm point groups on the other hand, bulk electric-dipole SHG is forbidden, and we instead assume the dominant contribution to be of bulk electric-quadrupole origin, consistent with the case for Sr$_{2}$IrO$_{4}$.\cite{Torch2015,Zhao2015} It is clear from Figs. \ref{fig:SHGplot} (e)\textendash(h) that none of these three proposed point groups can describe the RA-SHG data. On the other hand, by assuming bulk-electric quadrupole induced SHG from a centrosymmetric 4/m point group, we obtain excellent agreement with the data  as shown in Figs. \ref{fig:SHGplot} (a)\textendash(d) (We note that bulk magnetic dipole induced SHG from a 4/m point group does not qualitatively match the data). Any sub-group of 4/m (such as 2/m) fits the data equally well since it naturally contains all elements of the 4/m electric-quadrupole susceptibility tensor. Taken together with the diffraction data presented in Section III, these results suggest that Sr$_{3}$Ir$_{2}$O$_{7}$ crystallizes in a 2/m point group but is very close to being 4/m.

\begin{figure}
\includegraphics[scale=0.45]{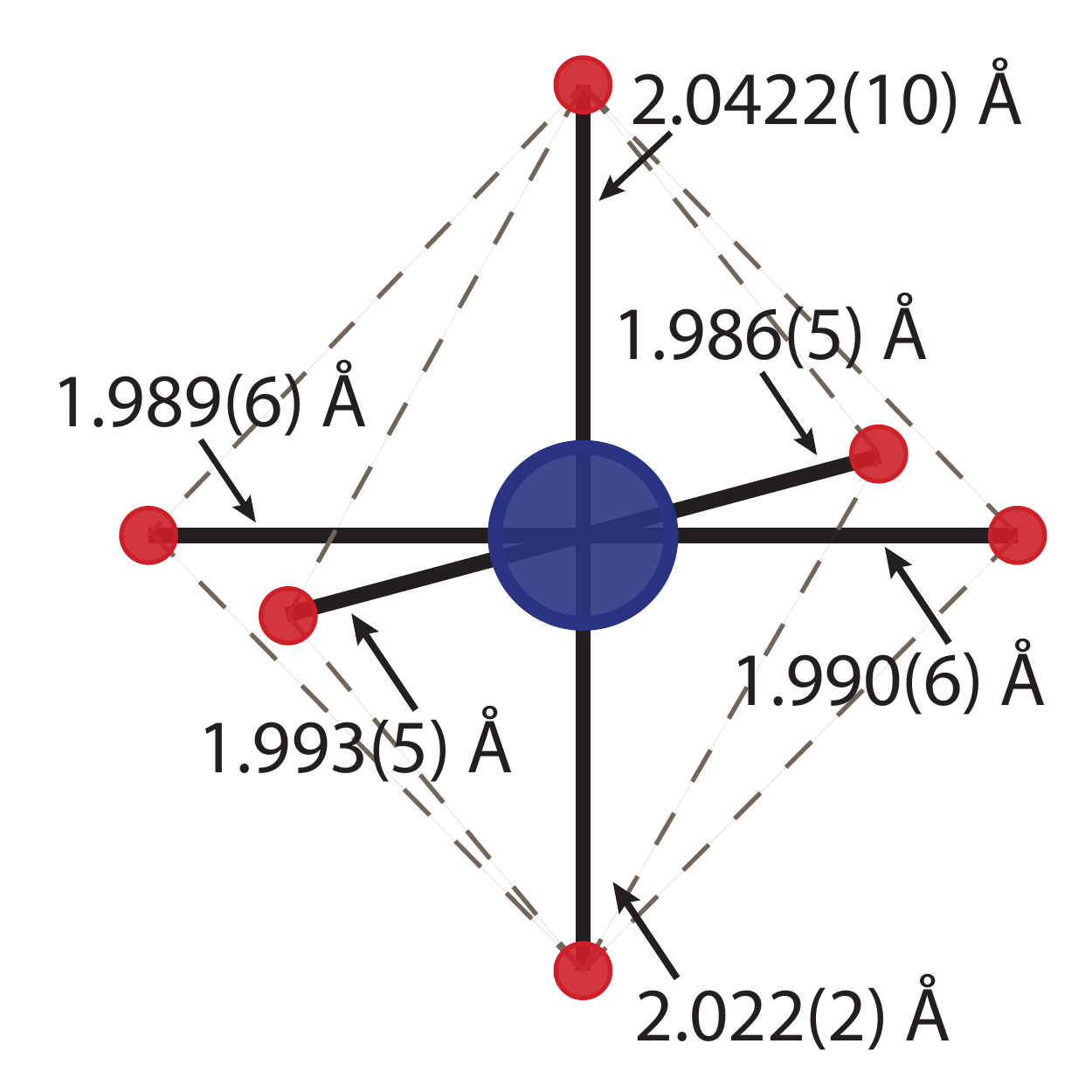}
\caption{\label{fig:oct} Refined \textit{T} = 300 K geometry of the IrO$_6$ octahedra in Sr$_{3}$Ir$_{2}$O$_{7}$ showing relevant bond lengths and their uncertainties.  Oxygen atoms are depicted in red, with the central iridium atom in blue.}  
\end{figure}

\section{Structural model of S\MakeLowercase{r}$_3$I\MakeLowercase{r}$_2$O$_7$}

With the RA-SHG analysis unambiguously ruling out the orthorhombic point group mmm associated with \textit{Ccce}, we exclude space group No.\ 68 as a possible solution and focus exclusively on No.\ 15 (\textit{C2/c}).  Tables \ref{tab:coords} and \ref{tab:coords100} list the complete refinement results obtained at $T = 300$ K and $T = 100$ K respectively.  The relative atomic positions reported correspond to the standard setting of space group \textit{C2/c} (No. 15) wherein the unique axis (associated with the single oblique angle $\beta$) is \textit{b}, and the long axis is now \textit{a}.  Atomic displacement factors were also refined with all $U_{ij}$ matrices passing a check for positive definiteness.  Each atomic site within the cell refined to be fully occupied within error.  Associated unit cell parameters can be found in Table \ref{tab:ucell}.  

To converge, fits require refining the structure as a twin and obtain final \textit{R1} values of 5.7\% (300 K) and 5.9\%(100 K).  Our choice of twin law is informed by the `accidental' pseudosymmetry of the unit cell.  Previously, the unit cell parameters comprising the basal plane (\textit{b} and \textit{c} for \textit{C2/c}) have been consistently reported as identical to within experimental precision and no measurable obliquity of the cell has been reported. As no tetragonal space groups produce a satisfactory solution, the conditions \textit{b} $\approx$ \textit{c}, $\beta$ $\approx$ 90\textsuperscript{$\circ$}, instead represent a effective tetragonal metric of the system.  The twin law which permits the refinement to converge is a symmetry operator of the tetragonal point group 4/mmm: a two-fold rotation about the [011] direction (basal plane diagonal).  In real space, for a small deviation from the condition $\beta$ = 90\textsuperscript{$\circ$}, this is tantamount to an altering of the rotational phasing of the octahedra (along the long axis) at the twin boundary.  The twin scale factor was refined to 0.497, very near the ideal `perfect twin' value of 0.5.  We note that because the twin law is not a symmetry operator of the point group of the individual's lattice (2/m), this is classified as twinning by pseudo-merohedry.\cite{CrStructureBook}

A representative oxygen octahedral cage surrounding each Ir site is depicted in Fig. \ref{fig:oct}, revealing only a slight ($\sim$2\%) tetragonal distortion in the apical direction (\textit{\^{a}}, for the standard setting used here).  In comparing this new model at 300 K with previous reports, two distinguishing attributes should be highlighted: first, the in-plane rotation angle of 11.5\textsuperscript{$\circ$} closely matches previous measurements.\cite{Cao2002}  Next, in contrast to \textit{Ccce}, the out-of-plane tilt (now permitted) refines to a value of 0.23\textsuperscript{$\circ$}.  At 100 K, the in-plane rotation angle increases to 11.8\textsuperscript{$\circ$} and the tilt to 0.33\textsuperscript{$\circ$}.  Representative projections of both features, alongside the full unit cell, are seen in Fig. \ref{fig:unitcell}.  The presence of a tilt angle representing only a $\sim$0.3\% deviation from the orthorhombic model with a nearly identical in-plane rotation at these temperatures is consistent with the fact that, by merit of diffraction data refinement alone, \textit{C2/c} and \textit{Ccce} describe the single crystal diffraction data equally well.

\begin{table}
\caption{\label{tab:ucell} Lattice Parameters for Sr$_{3}$Ir$_{2}$O$_{7}$ at measured temperatures. Space group \textit{C2/c}, \textit{Z} = 4.}
\begin{ruledtabular}
\begin{tabular}{cll}
 \hphantom{Parameters}  & \textit{T} = 300 K & \textit{T} = 100 K\\
 \hline
\textit{a} & 20.935(4) \AA 	& 20.917(3) \AA		\\
\textit{b} & 5.5185(13) \AA	& 5.5080(10) \AA	\\
\textit{c} & 5.5099(9) \AA	& 5.4995(7) \AA		\\
$\beta$ & 90.045(18)\textsuperscript{$\circ$}	& 90.069(15)\textsuperscript{$\circ$} \\
\textit{V} & 636.6(2) \AA\textsuperscript{3} & 633.60(17) \AA\textsuperscript{3} \\
	
\end{tabular}
\end{ruledtabular}
\end{table}


\section{Discussion}

As a separate metric aiding in the differentiation among possible space groups for Sr$_{3}$Ir$_{2}$O$_{7}$, density functional theory calculations were employed.  Atom positions in the two models for \textit{C2/c} and \textit{Ccce} were allowed to relax, subject to symmetry constraints, and the total energies of the resulting configurations were compared.  The energy associated with the monoclinic group  \textit{C2/c} was calculated to be 26 meV lower than that of \textit{Ccce}, supporting the notion that the activation of the octahedral tilt mode permits a slight reduction of the overall energetics. 

Based on the combination of neutron diffraction, RA-SHG data, and DFT analysis, the \textit{C2/c} model is a more complete structural solution as it also resolves previous anomalies in reported neutron diffraction data.  General reflection conditions for space group No.\ 15 (in the standard \textit{C2/c} setting) impose the condition H+L = 2\textit{n};\cite{IntTables} now the weak `violations' of the orthorhombic solution observed previously\cite{Dhital2014} are allowed reflections in \textit{C2/c}.  Structure factors corresponding to these reflections, calculated from the refined atomic positions in the monoclinic cell reported here, predict relative intensities of the order matching those observed in earlier triple-axis neutron studies.\cite{Dhital2014}  

The \textit{C2/c} model also offers a microscopic explanation of the anomalous presence of a net ferromagnetic moment in the basal plane of Sr$_{3}$Ir$_{2}$O$_{7}$ as seen in prior bulk susceptibility measurements.\cite{Cao2002,Hogan2015} The solution of the magnetic structure depicts antiferromagneticaly ordered moments aligned out-of-plane, along the \textit{\^{a}}-direction;\cite{Dhital2012,Fujiyama2012,Boseggia2012} however scattering would be unable to resolve small ($<1^{\circ}$) projections of the moment away from this axis.  Assuming a mechanism of spin-locking to octahedral orientation \cite{Jackeli2009} similar to that observed in Sr$_2$IrO$_4$, \cite{Feng2015, Boseggia2013} the ordered moment in the \textit{C2/c} setting would project a small component into the \textit{b-c} plane below $T_N$ = 280 K, resulting in a net ferromagnetic moment.   To quantify this, the tilt angle observed at $100$ K in conjunction with the reported ordered moment size of 0.36 $\mu_B$/Ir\cite{Dhital2012,Hogan2015} would imply an in-plane ferromagnetic moment of $\approx 1 \times 10^{-3}$ $\mu_B$/Ir, in relatively good agreement with \textit{b-c} plane magnetization data previously reported in this system.\cite{Nagai2007,Cao2002}

\section{Conclusions}

The measurements outlined here provide a comprehensive, multi-probe study arriving at the structural solution of the bilayer iridate system Sr$_{3}$Ir$_{2}$O$_{7}$.  The assignment of the monoclinic space group \textit{C2/c} (No.\ 15) readily accounts for previously reported rotations of the in-plane octahedra while at the same time enabling a subtle octahedral tilt mode distortion not resolved in previous studies of this system.  This tilt breaks the nominal orthorhombic symmetry, lowering the point group from mmm to 2/m as seen in the RA-SHG data, and permits scattering at previously observed Bragg violations of \textit{Ccce}.  Including this further distortion in the structural model is supported by DFT calculations, which demonstrate that such a tilt represents a lowering of the overall lattice energy. Our data provide the needed foundation for understanding how the lattice distorts and its subsequent role as this spin-orbit Mott system is driven toward the metallic state via doping or pressure.

\section{Acknowledgments}
This work was supported in part by NSF Award No. DMR-1505549 (S.D.W.), as well as by the \mbox{MRSEC} Program of the National Science Foundation under Award No. DMR 1121053 (T.H., L.B., C.VdW.).  Work at Caltech (L.Z., D.H.) was supported by ARO Grant W911NF-13-1-0059. Instrumentation for the SHG measurements was partially supported by ARO DURIP Award W911NF-13-1-0293 and by the Institute for Quantum Information and Matter, an NSF Physics Frontiers Center (PHY-1125565) with support of the Gordon and Betty Moore Foundation through Grant GBMF1250. C.B. acknowledges support from the Caltech WAVE Fellows program.  Work performed at the ORNL Spallation Neutron Source’s TOPAZ single-crystal diffractometer was supported by the Scientific User Facilities Division, Office of Basic Energy Sciences, US Department of Energy, under Contract No. DE-AC05-00OR22725 with UT-Battelle, LLC.  

\bibliography{BibTeX2}

\begin{thebibliography}{42}
\expandafter\ifx\csname natexlab\endcsname\relax\def\natexlab#1{#1}\fi
\expandafter\ifx\csname bibnamefont\endcsname\relax
  \def\bibnamefont#1{#1}\fi
\expandafter\ifx\csname bibfnamefont\endcsname\relax
  \def\bibfnamefont#1{#1}\fi
\expandafter\ifx\csname citenamefont\endcsname\relax
  \def\citenamefont#1{#1}\fi
\expandafter\ifx\csname url\endcsname\relax
  \def\url#1{\texttt{#1}}\fi
\expandafter\ifx\csname urlprefix\endcsname\relax\def\urlprefix{URL }\fi
\providecommand{\bibinfo}[2]{#2}
\providecommand{\eprint}[2][]{\url{#2}}

\bibitem[{\citenamefont{Moon et~al.}(2008)\citenamefont{Moon, Jin, Kim, Choi,
  Lee, Yu, Cao, Sumi, Funakubo, Bernhard et~al.}}]{Moon2008}
\bibinfo{author}{\bibfnamefont{S.~J.} \bibnamefont{Moon}},
  \bibinfo{author}{\bibfnamefont{H.}~\bibnamefont{Jin}},
  \bibinfo{author}{\bibfnamefont{K.~W.} \bibnamefont{Kim}},
  \bibinfo{author}{\bibfnamefont{W.~S.} \bibnamefont{Choi}},
  \bibinfo{author}{\bibfnamefont{Y.~S.} \bibnamefont{Lee}},
  \bibinfo{author}{\bibfnamefont{J.}~\bibnamefont{Yu}},
  \bibinfo{author}{\bibfnamefont{G.}~\bibnamefont{Cao}},
  \bibinfo{author}{\bibfnamefont{A.}~\bibnamefont{Sumi}},
  \bibinfo{author}{\bibfnamefont{H.}~\bibnamefont{Funakubo}},
  \bibinfo{author}{\bibfnamefont{C.}~\bibnamefont{Bernhard}},
  \bibnamefont{et~al.}, \bibinfo{journal}{Phys. Rev. Lett.}
  \textbf{\bibinfo{volume}{101}}, \bibinfo{pages}{226402}
  (\bibinfo{year}{2008}).

\bibitem[{\citenamefont{Kim et~al.}(2008)\citenamefont{Kim, Jin, Moon, Kim,
  Park, Leem, Yu, Noh, Kim, Oh et~al.}}]{BJKim2008}
\bibinfo{author}{\bibfnamefont{B.~J.} \bibnamefont{Kim}},
  \bibinfo{author}{\bibfnamefont{H.}~\bibnamefont{Jin}},
  \bibinfo{author}{\bibfnamefont{S.~J.} \bibnamefont{Moon}},
  \bibinfo{author}{\bibfnamefont{J.-Y.} \bibnamefont{Kim}},
  \bibinfo{author}{\bibfnamefont{B.-G.} \bibnamefont{Park}},
  \bibinfo{author}{\bibfnamefont{C.~S.} \bibnamefont{Leem}},
  \bibinfo{author}{\bibfnamefont{J.}~\bibnamefont{Yu}},
  \bibinfo{author}{\bibfnamefont{T.~W.} \bibnamefont{Noh}},
  \bibinfo{author}{\bibfnamefont{C.}~\bibnamefont{Kim}},
  \bibinfo{author}{\bibfnamefont{S.-J.} \bibnamefont{Oh}},
  \bibnamefont{et~al.}, \bibinfo{journal}{Phys. Rev. Lett.}
  \textbf{\bibinfo{volume}{101}}, \bibinfo{pages}{076402}
  (\bibinfo{year}{2008}).

\bibitem[{\citenamefont{Kim et~al.}(2009)\citenamefont{Kim, Ohsumi, Komesu,
  Sakai, Morita, Takagi, and Arima}}]{BJKim2009}
\bibinfo{author}{\bibfnamefont{B.~J.} \bibnamefont{Kim}},
  \bibinfo{author}{\bibfnamefont{H.}~\bibnamefont{Ohsumi}},
  \bibinfo{author}{\bibfnamefont{T.}~\bibnamefont{Komesu}},
  \bibinfo{author}{\bibfnamefont{S.}~\bibnamefont{Sakai}},
  \bibinfo{author}{\bibfnamefont{T.}~\bibnamefont{Morita}},
  \bibinfo{author}{\bibfnamefont{H.}~\bibnamefont{Takagi}}, \bibnamefont{and}
  \bibinfo{author}{\bibfnamefont{T.}~\bibnamefont{Arima}},
  \bibinfo{journal}{Science} \textbf{\bibinfo{volume}{323}},
  \bibinfo{pages}{1329} (\bibinfo{year}{2009}).

\bibitem[{\citenamefont{Jackeli and Khaliullin}(2009)}]{Jackeli2009}
\bibinfo{author}{\bibfnamefont{G.}~\bibnamefont{Jackeli}} \bibnamefont{and}
  \bibinfo{author}{\bibfnamefont{G.}~\bibnamefont{Khaliullin}},
  \bibinfo{journal}{Physical Review Letters} \textbf{\bibinfo{volume}{102}},
  \bibinfo{pages}{017205} (\bibinfo{year}{2009}).

\bibitem[{\citenamefont{Carter and Kee}(2013)}]{Carter2013}
\bibinfo{author}{\bibfnamefont{J.-M.} \bibnamefont{Carter}} \bibnamefont{and}
  \bibinfo{author}{\bibfnamefont{H.-Y.} \bibnamefont{Kee}},
  \bibinfo{journal}{Phys. Rev. B} \textbf{\bibinfo{volume}{87}},
  \bibinfo{pages}{014433} (\bibinfo{year}{2013}).

\bibitem[{\citenamefont{King et~al.}(2013)\citenamefont{King, Takayama, Tamai,
  Rozbicki, Walker, Shi, Patthey, Moore, Lu, Shen et~al.}}]{King2013}
\bibinfo{author}{\bibfnamefont{P.~D.~C.} \bibnamefont{King}},
  \bibinfo{author}{\bibfnamefont{T.}~\bibnamefont{Takayama}},
  \bibinfo{author}{\bibfnamefont{A.}~\bibnamefont{Tamai}},
  \bibinfo{author}{\bibfnamefont{E.}~\bibnamefont{Rozbicki}},
  \bibinfo{author}{\bibfnamefont{S.~M.} \bibnamefont{Walker}},
  \bibinfo{author}{\bibfnamefont{M.}~\bibnamefont{Shi}},
  \bibinfo{author}{\bibfnamefont{L.}~\bibnamefont{Patthey}},
  \bibinfo{author}{\bibfnamefont{R.~G.} \bibnamefont{Moore}},
  \bibinfo{author}{\bibfnamefont{D.}~\bibnamefont{Lu}},
  \bibinfo{author}{\bibfnamefont{K.~M.} \bibnamefont{Shen}},
  \bibnamefont{et~al.}, \bibinfo{journal}{Phys. Rev. B}
  \textbf{\bibinfo{volume}{87}}, \bibinfo{pages}{241106}
  (\bibinfo{year}{2013}).

\bibitem[{\citenamefont{Moreschini et~al.}(2014)\citenamefont{Moreschini,
  Moser, Ebrahimi, Dalla~Piazza, Kim, Boseggia, McMorrow, R\o{}nnow, Chang,
  Prabhakaran et~al.}}]{Grioni2014}
\bibinfo{author}{\bibfnamefont{L.}~\bibnamefont{Moreschini}},
  \bibinfo{author}{\bibfnamefont{S.}~\bibnamefont{Moser}},
  \bibinfo{author}{\bibfnamefont{A.}~\bibnamefont{Ebrahimi}},
  \bibinfo{author}{\bibfnamefont{B.}~\bibnamefont{Dalla~Piazza}},
  \bibinfo{author}{\bibfnamefont{K.~S.} \bibnamefont{Kim}},
  \bibinfo{author}{\bibfnamefont{S.}~\bibnamefont{Boseggia}},
  \bibinfo{author}{\bibfnamefont{D.~F.} \bibnamefont{McMorrow}},
  \bibinfo{author}{\bibfnamefont{H.~M.} \bibnamefont{R\o{}nnow}},
  \bibinfo{author}{\bibfnamefont{J.}~\bibnamefont{Chang}},
  \bibinfo{author}{\bibfnamefont{D.}~\bibnamefont{Prabhakaran}},
  \bibnamefont{et~al.}, \bibinfo{journal}{Phys. Rev. B}
  \textbf{\bibinfo{volume}{89}}, \bibinfo{pages}{201114}
  (\bibinfo{year}{2014}).

\bibitem[{\citenamefont{Liu et~al.}(2014)\citenamefont{Liu, Xu, Alidoust,
  Chang, Lin, Dhital, Khadka, Neupane, Belopolski, Landolt et~al.}}]{Liu2014}
\bibinfo{author}{\bibfnamefont{C.}~\bibnamefont{Liu}},
  \bibinfo{author}{\bibfnamefont{S.-Y.} \bibnamefont{Xu}},
  \bibinfo{author}{\bibfnamefont{N.}~\bibnamefont{Alidoust}},
  \bibinfo{author}{\bibfnamefont{T.-R.} \bibnamefont{Chang}},
  \bibinfo{author}{\bibfnamefont{H.}~\bibnamefont{Lin}},
  \bibinfo{author}{\bibfnamefont{C.}~\bibnamefont{Dhital}},
  \bibinfo{author}{\bibfnamefont{S.}~\bibnamefont{Khadka}},
  \bibinfo{author}{\bibfnamefont{M.}~\bibnamefont{Neupane}},
  \bibinfo{author}{\bibfnamefont{I.}~\bibnamefont{Belopolski}},
  \bibinfo{author}{\bibfnamefont{G.}~\bibnamefont{Landolt}},
  \bibnamefont{et~al.}, \bibinfo{journal}{Phys. Rev. B}
  \textbf{\bibinfo{volume}{90}}, \bibinfo{pages}{045127}
  (\bibinfo{year}{2014}).

\bibitem[{\citenamefont{Kim et~al.}(2012)\citenamefont{Kim, Choi, Kim,
  Mitchell, Jackeli, Daghofer, van~den Brink, Khaliullin, and
  Kim}}]{PhysRevLett.109.037204}
\bibinfo{author}{\bibfnamefont{J.~W.} \bibnamefont{Kim}},
  \bibinfo{author}{\bibfnamefont{Y.}~\bibnamefont{Choi}},
  \bibinfo{author}{\bibfnamefont{J.}~\bibnamefont{Kim}},
  \bibinfo{author}{\bibfnamefont{J.~F.} \bibnamefont{Mitchell}},
  \bibinfo{author}{\bibfnamefont{G.}~\bibnamefont{Jackeli}},
  \bibinfo{author}{\bibfnamefont{M.}~\bibnamefont{Daghofer}},
  \bibinfo{author}{\bibfnamefont{J.}~\bibnamefont{van~den Brink}},
  \bibinfo{author}{\bibfnamefont{G.}~\bibnamefont{Khaliullin}},
  \bibnamefont{and} \bibinfo{author}{\bibfnamefont{B.~J.} \bibnamefont{Kim}},
  \bibinfo{journal}{Phys. Rev. Lett.} \textbf{\bibinfo{volume}{109}},
  \bibinfo{pages}{037204} (\bibinfo{year}{2012}).

\bibitem[{\citenamefont{{Subramanian, M. A., Crawford, M. K.,
  Harlow}}(1994)}]{Subramanian94}
\bibinfo{author}{\bibfnamefont{R.~L.} \bibnamefont{{Subramanian, M. A.,
  Crawford, M. K., Harlow}}}, \bibinfo{journal}{Mater. Res. Bull.}
  \textbf{\bibinfo{volume}{29}}, \bibinfo{pages}{645} (\bibinfo{year}{1994}).

\bibitem[{\citenamefont{Boseggia et~al.}(2012)\citenamefont{Boseggia,
  Springell, Walker, Boothroyd, Prabhakaran, Wermeille, Bouchenoire, Collins,
  and McMorrow}}]{Boseggia2012}
\bibinfo{author}{\bibfnamefont{S.}~\bibnamefont{Boseggia}},
  \bibinfo{author}{\bibfnamefont{R.}~\bibnamefont{Springell}},
  \bibinfo{author}{\bibfnamefont{H.~C.} \bibnamefont{Walker}},
  \bibinfo{author}{\bibfnamefont{A.~T.} \bibnamefont{Boothroyd}},
  \bibinfo{author}{\bibfnamefont{D.}~\bibnamefont{Prabhakaran}},
  \bibinfo{author}{\bibfnamefont{D.}~\bibnamefont{Wermeille}},
  \bibinfo{author}{\bibfnamefont{L.}~\bibnamefont{Bouchenoire}},
  \bibinfo{author}{\bibfnamefont{S.~P.} \bibnamefont{Collins}},
  \bibnamefont{and} \bibinfo{author}{\bibfnamefont{D.~F.}
  \bibnamefont{McMorrow}}, \bibinfo{journal}{Phys. Rev. B}
  \textbf{\bibinfo{volume}{85}}, \bibinfo{pages}{184432}
  (\bibinfo{year}{2012}).

\bibitem[{\citenamefont{Cao et~al.}(2002)\citenamefont{Cao, Xin, Alexander,
  Crow, Schlottmann, Crawford, Harlow, and Marshall}}]{Cao2002}
\bibinfo{author}{\bibfnamefont{G.}~\bibnamefont{Cao}},
  \bibinfo{author}{\bibfnamefont{Y.}~\bibnamefont{Xin}},
  \bibinfo{author}{\bibfnamefont{C.}~\bibnamefont{Alexander}},
  \bibinfo{author}{\bibfnamefont{J.}~\bibnamefont{Crow}},
  \bibinfo{author}{\bibfnamefont{P.}~\bibnamefont{Schlottmann}},
  \bibinfo{author}{\bibfnamefont{M.}~\bibnamefont{Crawford}},
  \bibinfo{author}{\bibfnamefont{R.}~\bibnamefont{Harlow}}, \bibnamefont{and}
  \bibinfo{author}{\bibfnamefont{W.}~\bibnamefont{Marshall}},
  \bibinfo{journal}{Phys. Rev. B} \textbf{\bibinfo{volume}{66}},
  \bibinfo{pages}{214412} (\bibinfo{year}{2002}).

\bibitem[{\citenamefont{Matsuhata et~al.}(2004)\citenamefont{Matsuhata, Nagai,
  Yoshida, Hara, Ikeda, and Shirakawa}}]{Matsuhata2004}
\bibinfo{author}{\bibfnamefont{H.}~\bibnamefont{Matsuhata}},
  \bibinfo{author}{\bibfnamefont{I.}~\bibnamefont{Nagai}},
  \bibinfo{author}{\bibfnamefont{Y.}~\bibnamefont{Yoshida}},
  \bibinfo{author}{\bibfnamefont{S.}~\bibnamefont{Hara}},
  \bibinfo{author}{\bibfnamefont{S.-i.} \bibnamefont{Ikeda}}, \bibnamefont{and}
  \bibinfo{author}{\bibfnamefont{N.}~\bibnamefont{Shirakawa}},
  \bibinfo{journal}{J. Solid State Chem.} \textbf{\bibinfo{volume}{177}},
  \bibinfo{pages}{3776} (\bibinfo{year}{2004}).

\bibitem[{\citenamefont{Shaked et~al.}(2000)\citenamefont{Shaked, Jorgensen,
  Chmaissem, Ikeda, and Maeno}}]{Shaked2000}
\bibinfo{author}{\bibfnamefont{H.}~\bibnamefont{Shaked}},
  \bibinfo{author}{\bibfnamefont{J.~D.} \bibnamefont{Jorgensen}},
  \bibinfo{author}{\bibfnamefont{O.}~\bibnamefont{Chmaissem}},
  \bibinfo{author}{\bibfnamefont{S.}~\bibnamefont{Ikeda}}, \bibnamefont{and}
  \bibinfo{author}{\bibfnamefont{Y.}~\bibnamefont{Maeno}}, \bibinfo{journal}{J.
  Solid State Chem.} \textbf{\bibinfo{volume}{154}}, \bibinfo{pages}{361}
  (\bibinfo{year}{2000}).

\bibitem[{\citenamefont{Dhital et~al.}(2012)\citenamefont{Dhital, Khadka,
  Yamani, de~la Cruz, Hogan, Disseler, Pokharel, Lukas, Tian, Opeil
  et~al.}}]{Dhital2012}
\bibinfo{author}{\bibfnamefont{C.}~\bibnamefont{Dhital}},
  \bibinfo{author}{\bibfnamefont{S.}~\bibnamefont{Khadka}},
  \bibinfo{author}{\bibfnamefont{Z.}~\bibnamefont{Yamani}},
  \bibinfo{author}{\bibfnamefont{C.}~\bibnamefont{de~la Cruz}},
  \bibinfo{author}{\bibfnamefont{T.~C.} \bibnamefont{Hogan}},
  \bibinfo{author}{\bibfnamefont{S.~M.} \bibnamefont{Disseler}},
  \bibinfo{author}{\bibfnamefont{M.}~\bibnamefont{Pokharel}},
  \bibinfo{author}{\bibfnamefont{K.~C.} \bibnamefont{Lukas}},
  \bibinfo{author}{\bibfnamefont{W.}~\bibnamefont{Tian}},
  \bibinfo{author}{\bibfnamefont{C.~P.} \bibnamefont{Opeil}},
  \bibnamefont{et~al.}, \bibinfo{journal}{Phys. Rev. B}
  \textbf{\bibinfo{volume}{86}}, \bibinfo{pages}{100401}
  (\bibinfo{year}{2012}).

\bibitem[{\citenamefont{Hahn}(2006)}]{IntTables}
\bibinfo{editor}{\bibfnamefont{T.}~\bibnamefont{Hahn}}, ed.,
  \emph{\bibinfo{title}{International Tables for Crystallography}}, vol.
  \bibinfo{volume}{A, \textit{Space-group symmetry}}
  (\bibinfo{publisher}{Chester: International Union of Crystallography},
  \bibinfo{year}{2006}), \bibinfo{edition}{1st} ed.

\bibitem[{\citenamefont{Dhital et~al.}(2014)\citenamefont{Dhital, Hogan, Zhou,
  Chen, Ren, Pokharel, Okada, Heine, Tian, Yamani et~al.}}]{Dhital2014}
\bibinfo{author}{\bibfnamefont{C.}~\bibnamefont{Dhital}},
  \bibinfo{author}{\bibfnamefont{T.}~\bibnamefont{Hogan}},
  \bibinfo{author}{\bibfnamefont{W.}~\bibnamefont{Zhou}},
  \bibinfo{author}{\bibfnamefont{X.}~\bibnamefont{Chen}},
  \bibinfo{author}{\bibfnamefont{Z.}~\bibnamefont{Ren}},
  \bibinfo{author}{\bibfnamefont{M.}~\bibnamefont{Pokharel}},
  \bibinfo{author}{\bibfnamefont{Y.}~\bibnamefont{Okada}},
  \bibinfo{author}{\bibfnamefont{M.}~\bibnamefont{Heine}},
  \bibinfo{author}{\bibfnamefont{W.}~\bibnamefont{Tian}},
  \bibinfo{author}{\bibfnamefont{Z.}~\bibnamefont{Yamani}},
  \bibnamefont{et~al.}, \bibinfo{journal}{Nat. Commun.}
  \textbf{\bibinfo{volume}{5}}, \bibinfo{pages}{1} (\bibinfo{year}{2014}), ISSN
  \bibinfo{issn}{2041-1723}.

\bibitem[{\citenamefont{Ye et~al.}(2013)\citenamefont{Ye, Chi, Chakoumakos,
  Fernandez-Baca, Qi, and Cao}}]{Feng2013}
\bibinfo{author}{\bibfnamefont{F.}~\bibnamefont{Ye}},
  \bibinfo{author}{\bibfnamefont{S.}~\bibnamefont{Chi}},
  \bibinfo{author}{\bibfnamefont{B.~C.} \bibnamefont{Chakoumakos}},
  \bibinfo{author}{\bibfnamefont{J.~A.} \bibnamefont{Fernandez-Baca}},
  \bibinfo{author}{\bibfnamefont{T.}~\bibnamefont{Qi}}, \bibnamefont{and}
  \bibinfo{author}{\bibfnamefont{G.}~\bibnamefont{Cao}},
  \bibinfo{journal}{Phys. Rev. B} \textbf{\bibinfo{volume}{87}},
  \bibinfo{pages}{140406} (\bibinfo{year}{2013}).

\bibitem[{\citenamefont{Dhital et~al.}(2013)\citenamefont{Dhital, Hogan,
  Yamani, de~la Cruz, Chen, Khadka, Ren, and Wilson}}]{Dhital2013}
\bibinfo{author}{\bibfnamefont{C.}~\bibnamefont{Dhital}},
  \bibinfo{author}{\bibfnamefont{T.}~\bibnamefont{Hogan}},
  \bibinfo{author}{\bibfnamefont{Z.}~\bibnamefont{Yamani}},
  \bibinfo{author}{\bibfnamefont{C.}~\bibnamefont{de~la Cruz}},
  \bibinfo{author}{\bibfnamefont{X.}~\bibnamefont{Chen}},
  \bibinfo{author}{\bibfnamefont{S.}~\bibnamefont{Khadka}},
  \bibinfo{author}{\bibfnamefont{Z.}~\bibnamefont{Ren}}, \bibnamefont{and}
  \bibinfo{author}{\bibfnamefont{S.~D.} \bibnamefont{Wilson}},
  \bibinfo{journal}{Phys. Rev. B} \textbf{\bibinfo{volume}{87}},
  \bibinfo{pages}{144405} (\bibinfo{year}{2013}).

\bibitem[{\citenamefont{Torchinsky et~al.}(2015)\citenamefont{Torchinsky, Chu,
  Zhao, Perkins, Sizyuk, Qi, Cao, and Hsieh}}]{Torch2015}
\bibinfo{author}{\bibfnamefont{D.~H.} \bibnamefont{Torchinsky}},
  \bibinfo{author}{\bibfnamefont{H.}~\bibnamefont{Chu}},
  \bibinfo{author}{\bibfnamefont{L.}~\bibnamefont{Zhao}},
  \bibinfo{author}{\bibfnamefont{N.~B.} \bibnamefont{Perkins}},
  \bibinfo{author}{\bibfnamefont{Y.}~\bibnamefont{Sizyuk}},
  \bibinfo{author}{\bibfnamefont{T.}~\bibnamefont{Qi}},
  \bibinfo{author}{\bibfnamefont{G.}~\bibnamefont{Cao}}, \bibnamefont{and}
  \bibinfo{author}{\bibfnamefont{D.}~\bibnamefont{Hsieh}},
  \bibinfo{journal}{Phys. Rev. Lett.} \textbf{\bibinfo{volume}{114}},
  \bibinfo{pages}{096404} (\bibinfo{year}{2015}).

\bibitem[{\citenamefont{Ye et~al.}(2015)\citenamefont{Ye, Wang, Hoffmann, Wang,
  Chi, Matsuda, Chakoumakos, Fernandez-Baca, and Cao}}]{Feng2015}
\bibinfo{author}{\bibfnamefont{F.}~\bibnamefont{Ye}},
  \bibinfo{author}{\bibfnamefont{X.}~\bibnamefont{Wang}},
  \bibinfo{author}{\bibfnamefont{C.}~\bibnamefont{Hoffmann}},
  \bibinfo{author}{\bibfnamefont{J.}~\bibnamefont{Wang}},
  \bibinfo{author}{\bibfnamefont{S.}~\bibnamefont{Chi}},
  \bibinfo{author}{\bibfnamefont{M.}~\bibnamefont{Matsuda}},
  \bibinfo{author}{\bibfnamefont{B.~C.} \bibnamefont{Chakoumakos}},
  \bibinfo{author}{\bibfnamefont{J.~A.} \bibnamefont{Fernandez-Baca}},
  \bibnamefont{and} \bibinfo{author}{\bibfnamefont{G.}~\bibnamefont{Cao}},
  \bibinfo{journal}{Phys. Rev. B} \textbf{\bibinfo{volume}{92}},
  \bibinfo{pages}{201112} (\bibinfo{year}{2015}).

\bibitem[{\citenamefont{Zhao et~al.}(2014)\citenamefont{Zhao, Wang, Qi, Zeng,
  Hirai, Kong, Li, Park, Yuan, Jin et~al.}}]{Zhao2014}
\bibinfo{author}{\bibfnamefont{Z.}~\bibnamefont{Zhao}},
  \bibinfo{author}{\bibfnamefont{S.}~\bibnamefont{Wang}},
  \bibinfo{author}{\bibfnamefont{T.~F.} \bibnamefont{Qi}},
  \bibinfo{author}{\bibfnamefont{Q.}~\bibnamefont{Zeng}},
  \bibinfo{author}{\bibfnamefont{S.}~\bibnamefont{Hirai}},
  \bibinfo{author}{\bibfnamefont{P.~P.} \bibnamefont{Kong}},
  \bibinfo{author}{\bibfnamefont{L.}~\bibnamefont{Li}},
  \bibinfo{author}{\bibfnamefont{C.}~\bibnamefont{Park}},
  \bibinfo{author}{\bibfnamefont{S.~J.} \bibnamefont{Yuan}},
  \bibinfo{author}{\bibfnamefont{C.~Q.} \bibnamefont{Jin}},
  \bibnamefont{et~al.}, \bibinfo{journal}{J. Phys. Condens. Mat.}
  \textbf{\bibinfo{volume}{26}}, \bibinfo{pages}{215402}
  (\bibinfo{year}{2014}).

\bibitem[{\citenamefont{Hogan et~al.}(2015)\citenamefont{Hogan, Yamani, Walkup,
  Chen, Dally, Ward, Dean, Hill, Islam, Madhavan et~al.}}]{Hogan2015}
\bibinfo{author}{\bibfnamefont{T.}~\bibnamefont{Hogan}},
  \bibinfo{author}{\bibfnamefont{Z.}~\bibnamefont{Yamani}},
  \bibinfo{author}{\bibfnamefont{D.}~\bibnamefont{Walkup}},
  \bibinfo{author}{\bibfnamefont{X.}~\bibnamefont{Chen}},
  \bibinfo{author}{\bibfnamefont{R.}~\bibnamefont{Dally}},
  \bibinfo{author}{\bibfnamefont{T.~Z.} \bibnamefont{Ward}},
  \bibinfo{author}{\bibfnamefont{M.}~\bibnamefont{Dean}},
  \bibinfo{author}{\bibfnamefont{J.}~\bibnamefont{Hill}},
  \bibinfo{author}{\bibfnamefont{Z.}~\bibnamefont{Islam}},
  \bibinfo{author}{\bibfnamefont{V.}~\bibnamefont{Madhavan}},
  \bibnamefont{et~al.}, \bibinfo{journal}{Physical Review Letters}
  \textbf{\bibinfo{volume}{114}}, \bibinfo{pages}{257203}
  (\bibinfo{year}{2015}), \bibinfo{note}{\textit{Supplementary Information}}.

\bibitem[{\citenamefont{Zikovsky et~al.}(2011)\citenamefont{Zikovsky, Peterson,
  Wang, Frost, and Hoffmann}}]{crystalplan}
\bibinfo{author}{\bibfnamefont{J.}~\bibnamefont{Zikovsky}},
  \bibinfo{author}{\bibfnamefont{P.~F.} \bibnamefont{Peterson}},
  \bibinfo{author}{\bibfnamefont{X.~P.} \bibnamefont{Wang}},
  \bibinfo{author}{\bibfnamefont{M.}~\bibnamefont{Frost}}, \bibnamefont{and}
  \bibinfo{author}{\bibfnamefont{C.}~\bibnamefont{Hoffmann}},
  \bibinfo{journal}{J. Appl. Crystallogr.} \textbf{\bibinfo{volume}{44}},
  \bibinfo{pages}{418} (\bibinfo{year}{2011}).

\bibitem[{\citenamefont{Schultz et~al.}(1984)\citenamefont{Schultz, Srinivasan,
  Teller, Williams, and Lukehart}}]{anvred}
\bibinfo{author}{\bibfnamefont{A.~J.} \bibnamefont{Schultz}},
  \bibinfo{author}{\bibfnamefont{K.}~\bibnamefont{Srinivasan}},
  \bibinfo{author}{\bibfnamefont{R.~G.} \bibnamefont{Teller}},
  \bibinfo{author}{\bibfnamefont{J.~M.} \bibnamefont{Williams}},
  \bibnamefont{and} \bibinfo{author}{\bibfnamefont{C.~M.}
  \bibnamefont{Lukehart}}, \bibinfo{journal}{J. Am. Chem. Soc.}
  \textbf{\bibinfo{volume}{106}}, \bibinfo{pages}{999} (\bibinfo{year}{1984}).

\bibitem[{\citenamefont{Schultz et~al.}(2014)\citenamefont{Schultz,
  J{\o}rgensen, Wang, Mikkelson, Mikkelson, Lynch, Peterson, Green, and
  Hoffmann}}]{ellipint}
\bibinfo{author}{\bibfnamefont{A.~J.} \bibnamefont{Schultz}},
  \bibinfo{author}{\bibfnamefont{M.~R.~V.} \bibnamefont{J{\o}rgensen}},
  \bibinfo{author}{\bibfnamefont{X.}~\bibnamefont{Wang}},
  \bibinfo{author}{\bibfnamefont{R.~L.} \bibnamefont{Mikkelson}},
  \bibinfo{author}{\bibfnamefont{D.~J.} \bibnamefont{Mikkelson}},
  \bibinfo{author}{\bibfnamefont{V.~E.} \bibnamefont{Lynch}},
  \bibinfo{author}{\bibfnamefont{P.~F.} \bibnamefont{Peterson}},
  \bibinfo{author}{\bibfnamefont{M.~L.} \bibnamefont{Green}}, \bibnamefont{and}
  \bibinfo{author}{\bibfnamefont{C.~M.} \bibnamefont{Hoffmann}},
  \bibinfo{journal}{J. Appl. Crystallogr.} \textbf{\bibinfo{volume}{47}},
  \bibinfo{pages}{915} (\bibinfo{year}{2014}).

\bibitem[{\citenamefont{Sheldrick}(2008)}]{shelx}
\bibinfo{author}{\bibfnamefont{G.~M.} \bibnamefont{Sheldrick}},
  \bibinfo{journal}{Acta Crystallogr. A} \textbf{\bibinfo{volume}{64}},
  \bibinfo{pages}{112} (\bibinfo{year}{2008}).

\bibitem[{\citenamefont{Bl\"ochl}(1994)}]{PAW}
\bibinfo{author}{\bibfnamefont{P.~E.} \bibnamefont{Bl\"ochl}},
  \bibinfo{journal}{Phys. Rev. B} \textbf{\bibinfo{volume}{50}},
  \bibinfo{pages}{17953} (\bibinfo{year}{1994}).

\bibitem[{\citenamefont{Kresse and Joubert}(1999)}]{KresseVASP1}
\bibinfo{author}{\bibfnamefont{G.}~\bibnamefont{Kresse}} \bibnamefont{and}
  \bibinfo{author}{\bibfnamefont{D.}~\bibnamefont{Joubert}},
  \bibinfo{journal}{Phys. Rev. B} \textbf{\bibinfo{volume}{59}},
  \bibinfo{pages}{1758} (\bibinfo{year}{1999}).

\bibitem[{\citenamefont{Kresse and Furthmüller}(1996)}]{KresseVASP2}
\bibinfo{author}{\bibfnamefont{G.}~\bibnamefont{Kresse}} \bibnamefont{and}
  \bibinfo{author}{\bibfnamefont{J.}~\bibnamefont{Furthmüller}},
  \bibinfo{journal}{Comp. Mater. Sci.} \textbf{\bibinfo{volume}{6}},
  \bibinfo{pages}{15 } (\bibinfo{year}{1996}), ISSN \bibinfo{issn}{0927-0256}.

\bibitem[{\citenamefont{Perdew et~al.}(1996)\citenamefont{Perdew, Burke, and
  Ernzerhof}}]{PBE}
\bibinfo{author}{\bibfnamefont{J.~P.} \bibnamefont{Perdew}},
  \bibinfo{author}{\bibfnamefont{K.}~\bibnamefont{Burke}}, \bibnamefont{and}
  \bibinfo{author}{\bibfnamefont{M.}~\bibnamefont{Ernzerhof}},
  \bibinfo{journal}{Phys. Rev. Lett.} \textbf{\bibinfo{volume}{77}},
  \bibinfo{pages}{3865} (\bibinfo{year}{1996}).

\bibitem[{\citenamefont{Becke and Johnson}(2005)}]{BJ2005}
\bibinfo{author}{\bibfnamefont{A.~D.} \bibnamefont{Becke}} \bibnamefont{and}
  \bibinfo{author}{\bibfnamefont{E.~R.} \bibnamefont{Johnson}},
  \bibinfo{journal}{J. Chem. Phys.} \textbf{\bibinfo{volume}{123}},
  \bibinfo{eid}{154101} (\bibinfo{year}{2005}).

\bibitem[{\citenamefont{Torchinsky et~al.}(2014)\citenamefont{Torchinsky, Chu,
  Qi, Cao, and Hsieh}}]{Torch2014}
\bibinfo{author}{\bibfnamefont{D.~H.} \bibnamefont{Torchinsky}},
  \bibinfo{author}{\bibfnamefont{H.}~\bibnamefont{Chu}},
  \bibinfo{author}{\bibfnamefont{T.}~\bibnamefont{Qi}},
  \bibinfo{author}{\bibfnamefont{G.}~\bibnamefont{Cao}}, \bibnamefont{and}
  \bibinfo{author}{\bibfnamefont{D.}~\bibnamefont{Hsieh}},
  \bibinfo{journal}{Review of Scientific Instruments}
  \textbf{\bibinfo{volume}{85}}, \bibinfo{eid}{083102} (\bibinfo{year}{2014}).

\bibitem[{\citenamefont{Aroyo et~al.}(2011)\citenamefont{Aroyo, Perez-Mato,
  Orobengoa, Tasci, De~La~Flor, and Kirov}}]{Bilbao1}
\bibinfo{author}{\bibfnamefont{M.}~\bibnamefont{Aroyo}},
  \bibinfo{author}{\bibfnamefont{J.}~\bibnamefont{Perez-Mato}},
  \bibinfo{author}{\bibfnamefont{D.}~\bibnamefont{Orobengoa}},
  \bibinfo{author}{\bibfnamefont{E.}~\bibnamefont{Tasci}},
  \bibinfo{author}{\bibfnamefont{G.}~\bibnamefont{De~La~Flor}},
  \bibnamefont{and} \bibinfo{author}{\bibfnamefont{A.}~\bibnamefont{Kirov}},
  \bibinfo{journal}{Bulg. Chem. Commun.} \textbf{\bibinfo{volume}{43}},
  \bibinfo{pages}{183} (\bibinfo{year}{2011}).

\bibitem[{\citenamefont{Aroyo et~al.}(2009)\citenamefont{Aroyo, Perez-Mato,
  Capillas, and et~al.}}]{Bilbao2}
\bibinfo{author}{\bibfnamefont{M.}~\bibnamefont{Aroyo}},
  \bibinfo{author}{\bibfnamefont{J.}~\bibnamefont{Perez-Mato}},
  \bibinfo{author}{\bibfnamefont{C.}~\bibnamefont{Capillas}}, \bibnamefont{and}
  \bibinfo{author}{\bibnamefont{et~al.}}, \bibinfo{journal}{Z. Kristallog.}
  \textbf{\bibinfo{volume}{221}}, \bibinfo{pages}{15} (\bibinfo{year}{2009}).

\bibitem[{\citenamefont{Aroyo et~al.}(2006)\citenamefont{Aroyo, Kirov,
  Capillas, Perez-Mato, and Wondratschek}}]{Bilbao3}
\bibinfo{author}{\bibfnamefont{M.~I.} \bibnamefont{Aroyo}},
  \bibinfo{author}{\bibfnamefont{A.}~\bibnamefont{Kirov}},
  \bibinfo{author}{\bibfnamefont{C.}~\bibnamefont{Capillas}},
  \bibinfo{author}{\bibfnamefont{J.~M.} \bibnamefont{Perez-Mato}},
  \bibnamefont{and}
  \bibinfo{author}{\bibfnamefont{H.}~\bibnamefont{Wondratschek}},
  \bibinfo{journal}{Acta Crystallogr. A} \textbf{\bibinfo{volume}{62}},
  \bibinfo{pages}{115} (\bibinfo{year}{2006}).

\bibitem[{\citenamefont{Ivantchev et~al.}(2000)\citenamefont{Ivantchev,
  Kroumova, Madariaga, P{\'{e}}rez-Mato, and Aroyo}}]{subgroup}
\bibinfo{author}{\bibfnamefont{S.}~\bibnamefont{Ivantchev}},
  \bibinfo{author}{\bibfnamefont{E.}~\bibnamefont{Kroumova}},
  \bibinfo{author}{\bibfnamefont{G.}~\bibnamefont{Madariaga}},
  \bibinfo{author}{\bibfnamefont{J.~M.} \bibnamefont{P{\'{e}}rez-Mato}},
  \bibnamefont{and} \bibinfo{author}{\bibfnamefont{M.~I.} \bibnamefont{Aroyo}},
  \bibinfo{journal}{J. Appl. Crystallogr.} \textbf{\bibinfo{volume}{33}},
  \bibinfo{pages}{1190} (\bibinfo{year}{2000}).

\bibitem[{\citenamefont{Zhao et~al.}(2015)\citenamefont{Zhao, Torchinsky, Chu,
  Ivanov, Lifshitz, Flint, Qi, Cao, and Hsieh}}]{Zhao2015}
\bibinfo{author}{\bibfnamefont{L.}~\bibnamefont{Zhao}},
  \bibinfo{author}{\bibfnamefont{D.~H.} \bibnamefont{Torchinsky}},
  \bibinfo{author}{\bibfnamefont{H.}~\bibnamefont{Chu}},
  \bibinfo{author}{\bibfnamefont{V.}~\bibnamefont{Ivanov}},
  \bibinfo{author}{\bibfnamefont{R.}~\bibnamefont{Lifshitz}},
  \bibinfo{author}{\bibfnamefont{R.}~\bibnamefont{Flint}},
  \bibinfo{author}{\bibfnamefont{T.}~\bibnamefont{Qi}},
  \bibinfo{author}{\bibfnamefont{G.}~\bibnamefont{Cao}}, \bibnamefont{and}
  \bibinfo{author}{\bibfnamefont{D.}~\bibnamefont{Hsieh}},
  \bibinfo{journal}{Nat. Phys.}  (\bibinfo{year}{2015}),
  \bibinfo{note}{\textit{advance online publication}}.

\bibitem[{\citenamefont{Blake et~al.}(2009)\citenamefont{Blake, Clegg, Cole,
  Evans, Main, Parsons, and Watkin}}]{CrStructureBook}
\bibinfo{author}{\bibfnamefont{A.~J.} \bibnamefont{Blake}},
  \bibinfo{author}{\bibfnamefont{W.}~\bibnamefont{Clegg}},
  \bibinfo{author}{\bibfnamefont{J.~M.} \bibnamefont{Cole}},
  \bibinfo{author}{\bibfnamefont{J.~S.~O.} \bibnamefont{Evans}},
  \bibinfo{author}{\bibfnamefont{P.}~\bibnamefont{Main}},
  \bibinfo{author}{\bibfnamefont{S.}~\bibnamefont{Parsons}}, \bibnamefont{and}
  \bibinfo{author}{\bibfnamefont{D.~J.} \bibnamefont{Watkin}},
  \emph{\bibinfo{title}{Crystal Structure Analysis: Principles and Practice}},
  International Union of Crystallogrphy Book Series (\bibinfo{publisher}{Oxford
  University Press}, \bibinfo{year}{2009}), \bibinfo{edition}{2nd} ed.

\bibitem[{\citenamefont{Fujiyama et~al.}(2012)\citenamefont{Fujiyama, Ohashi,
  Ohsumi, Sugimoto, Takayama, Komesu, Takata, Arima, and
  Takagi}}]{Fujiyama2012}
\bibinfo{author}{\bibfnamefont{S.}~\bibnamefont{Fujiyama}},
  \bibinfo{author}{\bibfnamefont{K.}~\bibnamefont{Ohashi}},
  \bibinfo{author}{\bibfnamefont{H.}~\bibnamefont{Ohsumi}},
  \bibinfo{author}{\bibfnamefont{K.}~\bibnamefont{Sugimoto}},
  \bibinfo{author}{\bibfnamefont{T.}~\bibnamefont{Takayama}},
  \bibinfo{author}{\bibfnamefont{T.}~\bibnamefont{Komesu}},
  \bibinfo{author}{\bibfnamefont{M.}~\bibnamefont{Takata}},
  \bibinfo{author}{\bibfnamefont{T.}~\bibnamefont{Arima}}, \bibnamefont{and}
  \bibinfo{author}{\bibfnamefont{H.}~\bibnamefont{Takagi}},
  \bibinfo{journal}{Phys. Rev. B} \textbf{\bibinfo{volume}{86}},
  \bibinfo{pages}{174414} (\bibinfo{year}{2012}).

\bibitem[{\citenamefont{Boseggia et~al.}(2013)\citenamefont{Boseggia, Walker,
  Vale, Springell, Feng, Perry, Sala, Rønnow, Collins, and
  McMorrow}}]{Boseggia2013}
\bibinfo{author}{\bibfnamefont{S.}~\bibnamefont{Boseggia}},
  \bibinfo{author}{\bibfnamefont{H.~C.} \bibnamefont{Walker}},
  \bibinfo{author}{\bibfnamefont{J.}~\bibnamefont{Vale}},
  \bibinfo{author}{\bibfnamefont{R.}~\bibnamefont{Springell}},
  \bibinfo{author}{\bibfnamefont{Z.}~\bibnamefont{Feng}},
  \bibinfo{author}{\bibfnamefont{R.~S.} \bibnamefont{Perry}},
  \bibinfo{author}{\bibfnamefont{M.~M.} \bibnamefont{Sala}},
  \bibinfo{author}{\bibfnamefont{H.~M.} \bibnamefont{Rønnow}},
  \bibinfo{author}{\bibfnamefont{S.~P.} \bibnamefont{Collins}},
  \bibnamefont{and} \bibinfo{author}{\bibfnamefont{D.~F.}
  \bibnamefont{McMorrow}}, \bibinfo{journal}{Journal of Physics: Condensed
  Matter} \textbf{\bibinfo{volume}{25}}, \bibinfo{pages}{422202}
  (\bibinfo{year}{2013}).

\bibitem[{\citenamefont{Nagai et~al.}(2007)\citenamefont{Nagai, Yoshida, Ikeda,
  Matsuhata, Kito, and Kosaka}}]{Nagai2007}
\bibinfo{author}{\bibfnamefont{I.}~\bibnamefont{Nagai}},
  \bibinfo{author}{\bibfnamefont{Y.}~\bibnamefont{Yoshida}},
  \bibinfo{author}{\bibfnamefont{S.~I.} \bibnamefont{Ikeda}},
  \bibinfo{author}{\bibfnamefont{H.}~\bibnamefont{Matsuhata}},
  \bibinfo{author}{\bibfnamefont{H.}~\bibnamefont{Kito}}, \bibnamefont{and}
  \bibinfo{author}{\bibfnamefont{M.}~\bibnamefont{Kosaka}},
  \bibinfo{journal}{J. Phys. Condens. Mat.} \textbf{\bibinfo{volume}{19}},
  \bibinfo{pages}{136214} (\bibinfo{year}{2007}).

\end{thebibliography}

\end{document}